\documentclass[useAMS,usenatbib,a4]{mn2e}

\usepackage[utf8]{inputenc}
\usepackage{amsmath}
\usepackage{amssymb}
\usepackage{graphicx}
\usepackage{hyperref}
\usepackage{setspace}
\usepackage{graphics}
\usepackage{longtable}
\usepackage{indentfirst}
\usepackage{verbatim}
\usepackage{array}
\usepackage{rotating}
\usepackage{ifthen}
\usepackage{appendix}

\usepackage{multicol}

\newcommand{\sgn}{\mathop{\mathrm{sgn}}}

\title[Rotochemical heating with neutron superfluid gap models]{Rotochemical heating of millisecond and classical pulsars with anisotropic and density-dependent superfluid gap models}
\author[Nicol\'as Gonz\'alez-Jim\'enez, Cristobal Petrovich \& Andreas Reisenegger]{Nicol\'as Gonz\'alez-Jim\'enez$^{1,2}$, Cristobal Petrovich$^{1,3}$ \& Andreas Reisenegger$^{1}$\\
1: Instituto de Astrof\'isica, Facultad de F\'isica, Pontificia Universidad Cat\'olica de Chile, Av. Vicuña Mackenna 4860,\\ Macul, 7820436, Santiago, Chile \\
2: Argelander Institut für Astronomie, Universität Bonn, Auf dem Hügel 71, 53121, Bonn, Germany \\ E-mail: ngonzalez@astro.uni-bonn.de \\
3: Department of Astrophysical Sciences, Princeton University, 4 Ivy Ln, NJ 08544, Princeton, United States \\
}

\begin{document}
\maketitle
\begin{abstract}
When a rotating neutron star loses angular momentum, the progressive reduction of the centrifugal force makes it contract. This perturbs each fluid element, raising the local pressure and originating deviations from beta equilibrium, inducing reactions that release heat (``rotochemical heating''). This effect has previously been studied by \cite{F-R} for non-superfluid neutron stars and by \cite{petroka} for superfluid millisecond pulsars. Both studies found that pulsars reach a quasi-steady state in which the compression driving the matter out of beta equilibrium is balanced by the reactions trying to restore the equilibrium.\\ 
We extend previous studies by considering the effect of density-dependence and anisotropy of the superfluid energy gaps, for the case in which the dominant reactions are the modified Urca processes, the protons are non-superconducting,  and the neutron superfluidity is parametrized by models proposed in the literature. By comparing our predictions with the surface temperature of the millisecond pulsar PSR J0437-4715 and upper limits for twenty-one classical pulsars, we find the millisecond pulsar can be only explained by the models with the effectively largest energy gaps (type B models), the classical pulsars require with the gap models that vanish for some angle (type C) and two different envelope compositions. Thus, no single model for neutron superfluidity can simultaneously account for the thermal emission of all available observations of non-accreting neutron stars, possibly due to our neglect of proton superconductivity.

\end{abstract}

\section{Introduction}
\label{sec:intro}

A neutron star loses the thermal energy with which it was born, initially through neutrino emission, and after an age of $\sim10^{5}$ yr through photon emission \citep{yak01}. However, for late stages of the thermal evolution, several authors have proposed models for heating the matter due to different mechanisms, like vortex creep \citep{alpar} and rotochemical heating (\citealt{reis95}; improved later by \citealt{F-R}). \cite{denix} studied different mechanisms that can heat neutron stars, without considering the effects of finite energy gaps on the reaction rates in the neutron star interior. Among the mechanisms proposed, only the two mentioned above appear to account for the relatively high temperature inferred for the millisecond pulsar PSR J0437-4715 (\citealp{kargaltsev04}; \citealp{Durant}). It was also found that, in the case of vortex creep, the expected surface temperatures of several old classical pulsars ($\sim 10^{6-8}\mathrm{yr}$) would not lie much below current observational upper limits. 
Thus,
 if these limits could be lowered, they would either confirm or rule out 
vortex creep as the main heating mechanism, in the latter case indirectly confirming rotochemical heating as the only viable alternative proposed so far.

Rotochemical heating has its origin in deviations from beta equilibrium. As a neutron star reduces its rotation rate, the centrifugal force diminishes. This makes the star contract, perturbing each fluid element, raising the local pressure and originating deviations from beta equilibrium. The resulting non-equilibrium reactions release the energy stored in the chemical imbalance, which is partly emitted as neutrinos and partly converted into internal heat.

The most important prediction associated with rotochemical heating
is that, if the spin-down timescale is substantially longer than
any other timescale involved (with the exception of
magnetic field decay), the star arrives at a
quasi-steady state, where the rate at which neutrino reactions restore the equilibrium is the same
at which the spin-down modifies the equilibrium concentrations. In this state, the temperature depends
only on the current, slowly changing value of $\Omega
\dot{\Omega}$, the product of the angular velocity and its time
derivative, and not on its previous history (\citealt{reis95}; \citealt{F-R}). This allows a simple way
to constrain the physics involved in theoretical models, once the
spin parameters and observed surface temperature of a millisecond pulsar (MSP) are known.

Another relevant ingredient of NS cooling theory is the effect of superfluidity. It is well accepted that some particles in the interior of the NSs are in the superfluid state. This was first predicted by \cite{migdal60}, who proposed the NSs as good candidates to be macroscopic superfluid systems. From that prediction until now, the presence of neutron and proton superfluid phases has been studied to explain many properties of NSs.

The standard formalism to describe fermion superfluidity is the BCS theory of superconductivity (\citealt{BCS}). One of its main predictions is the existence of an energy gap in the quasi-particle density of states, located at the Fermi level. In normal matter, fermion states are filled up to the Fermi energy, and there is a finite density of states at the Fermi level. But in a BCS superfluid below a certain critical temperature $T_c$, the density of states acquires a gap of width $2\delta$ between the occupied and unoccupied states. Since the amplitude of this gap is very model-dependent, we need accurate quantitative theoretical predictions of its properties in order to understand the physics involved. Critical temperatures $T_c$ of neutrons and protons have been calculated by many authors, as reviewed by \cite{lombardo}. The results are very sensitive to the strong interaction models and many-body theories employed.



Following preliminary estimates by \cite{reis97}, \citet*[][hereafter PR10]{petroka}  were the first to model the thermal evolution of neutron stars with rotochemical heating, including the effects of superfluid energy gaps. Restricting themselves to the simplified case of spatially uniform and isotropic gaps for neutrons and protons, they were able to account for the surface temperature of MSP PSR J0437-4715, which is somewhat higher (a factor of $\sim 2$) than expected in non-superfluid models.

In this work, we go beyond those simple models by considering various density-dependent and anisotropic neutron energy gaps that have been proposed on the basis of theoretical models, but ignoring proton superconductivity. Restricting ourselves to modified Urca (Murca) reactions, we include these effects in the general calculation scheme used by \cite{F-R} and PR10 in order to follow the thermal evolution of neutron stars and verify if rotochemical heating can account for their observed temperatures (and upper limits).


The structure of this work is the following. In \S\ref{sec:theo} we review the theoretical framework of rotochemical heating without and with superfluidity. In \S\ref{sec:results}, we study the features of rotochemical heating in two different regimes, corresponding to millisecond pulsars and classical pulsars. We contrast our results against the surface temperature of PSR J0437-4715 and the upper limit of twenty one younger pulsars. We present our conclusions in \S\ref{sec:conclusions}. Finally in Appendix \ref{sec:appendix1} we describe the numerical approach to calculate the density-dependence of the superfluid gaps in our code, and in Appendix \ref{sec:appendix2} we describe our method to deal with the superfluid anisotropy of the models in order to compute the reduction factors for the emissivities and net reaction rates.
\section{Theoretical Framework}
\label{sec:theo}
\subsection{Thermal evolution with rotochemical heating}
\begin{table*}
\centering
\begin{tabular}{|c||c|c|c|c|}
\hline
 & Superfluidity type& $\lambda$ & $F(\vartheta)$ & $\Delta(0)/kT_c$\\
\hline
A & $^1S_0$ & 1 & 1&1.764\\
B & $^3P_2(|m_j|=0)$ & $1/2$ & $1+\cos^2\vartheta $ & 1.188 \\
C & $^3P_2(|m_j|=2)$ & $3/2$ & $\sin^2\vartheta $ & 2.03 \\
\hline
\end{tabular}
\caption{Parameters for the three standard types of superfluidity in neutron stars.}
\label{tabla1} 
\end{table*}
In this section, we present the most relevant equations used in this work. For the full theoretical derivations, see \cite{F-R} and PR10.
Throughout this work, we consider $npe\mu$ matter in the core of the star, i.e., neutrons, protons, electrons, and muons.

Due to the long timescales involved, we assume that thermal relaxation 
from an initial non-uniform internal temperature profile has already occurred, hence the redshifted internal temperature is uniform
\citep{glend97} and has the form:
\begin{eqnarray}
\label{eq:T_int}
T_\infty = T(r)e^{\Phi(r)},
\end{eqnarray} 
where $g_{tt}= -e^{2\Phi}$ is the time component of the metric of a non-rotating reference star, of which $r$
is the radial spherical coordinate. This condition is valid for all but the youngest neutron stars ($t<10^{3} \; \mathrm{years}$), because their evolutionary timescale is much longer than the heat diffusion time \citep{reis95}. In some cases, we show results even for the earliest stages for illustrative purposes, but we compare to data only in the regime where our approximation is accurate. 
The evolution of the internal temperature is given by the thermal balance equation \citep{thorne77}, which for an isothermal interior is given by
\begin{eqnarray}
\label{eq:dot_T}
\dot{T}_\infty = \frac{1}{C}\left[ L^{\infty}_H - L^{\infty}_\nu - L^{\infty}_\gamma \right],
\end{eqnarray}
where $C$ is the total heat capacity of the star, $L^{\infty}_H$ is the total power released by heating
mechanisms, $L^{\infty}_\nu$ is the total power emitted as neutrinos due to Urca reactions (in our case only Murca reactions) and Cooper pair-breaking and pair-formation processes (PBF), and $L^{\infty}_\gamma$ is the photon luminosity. The quantities $C,\;L^{\infty}_H,\;L^{\infty}_\nu$ and $L^{\infty}_\gamma$ remain unchanged from the definition in PR10.

Another relevant variable is the departure from the beta equilibrium due to the compression effect. This departure can be
quantified by the chemical imbalances \citep{haen92}:
\begin{eqnarray}
\label{eq:eta1}
\eta_{npl}=\delta \mu_n - \delta \mu_p - \delta \mu_l
\end{eqnarray}
where $l=e,\mu$, and $\delta \mu_i = \mu_i - \mu_i^{eq}$ are the deviation from
the chemical potential equilibrium of all species $i$, which include neutrons (n), protons (p), electrons (e), and muons ($\mu$). For the same reason that we consider a uniform temperature, we assume a uniform 
redshifted chemical potential deviation throughout the core,
\begin{eqnarray}
\label{eq:deltamu}
\delta \mu_{i}^{\infty} = \delta \mu_{i}(r) e^{\Phi(r)}.
\end{eqnarray}
We write the total energy dissipation rate as
\begin{eqnarray}
 L^{\infty}_H= \eta_{npe}^{\infty}\Delta\tilde\Gamma_{npe} + \eta_{np\mu}^{\infty}\Delta\tilde\Gamma_{np\mu},
\end{eqnarray}
where $\Delta\tilde\Gamma_{npl}=\tilde\Gamma_{n\rightarrow pl} - \tilde\Gamma_{pl\rightarrow n}$ is the net reaction rate  
integrated over the core (indicated with the tilde) involving the lepton $l$, a function of $\eta$ and $T$.
Finally, the evolution of the redshifted chemical imbalances is given by
\begin{eqnarray}
\label{eq:etaepunto_evol}
\dot\eta_{npe}^{\infty}= - Z_{npe}\Delta\tilde\Gamma_{npe}  - Z_{np}\Delta\tilde\Gamma_{np\mu} + 2W_{npe}\Omega\dot\Omega,\\
\label{eq:etamupunto_evol}
\dot\eta_{np\mu}^{\infty}= - Z_{np}\Delta\tilde\Gamma_{npe}  - Z_{np\mu}\Delta\tilde\Gamma_{np\mu} + 2W_{np\mu}\Omega\dot\Omega,
\end{eqnarray}
where the terms $Z_{np},\;Z_{npl}$, and $W_{npl}$ (with $l=e,\mu$) are constants that depend on the stellar structure and are kept unchanged with respect to their definition in \cite{reis06}.
Equations (\ref{eq:dot_T}), (\ref{eq:etaepunto_evol}), and (\ref{eq:etamupunto_evol}) give a complete description of the thermal evolution and the chemical imbalances of a neutron star with rotochemical heating and $npe\mu$ matter.

\subsection{Effects of superfluidity}
\label{sec:super}
The inclusion of superfluidity directly affects several quantities of the star, as it was explained in PR10. In this section, we describe the quantities relevant to our work.

The main effect comes from the energy gap $\delta$, which strongly influences the processes associated with the particles near the Fermi surface, such as the heat capacity and neutrino emission \citep{yak01}. In the superfluid state, the energy of a (quasi-)particle relative to the Fermi energy is:
\begin{eqnarray}
\label{eq:dispertion}
E=\pm \sqrt{(\epsilon-\mu)^2+\delta^2},
\end{eqnarray}
where $\epsilon (\mathbf{p})$ is the energy of a normal particle state of momentum $\mathbf{p}$, and $\mu$ is the chemical potential, which in the low-temperature limit becomes equal to the Fermi energy.
In the core of a neutron star, neutrons are believed to form Cooper pairs due to their interaction in the triplet $^{3}P_2$ state, while protons form singlet $^{1}P_0$ pairs. The Cooper pairing appears as a result of the attraction between particles with anti-parallel momenta. Its effect is most pronounced around the Fermi surface. 
Following \cite{yak01}, we parametrize the energy gap as:
\begin{eqnarray}
\label{eq:deltagap}
 \delta^{2} = \Delta(T)^2F(\vartheta),
\end{eqnarray}
where $\Delta(T)$ is the amplitude that contains the temperature dependence of the gap, and $F(\vartheta)$  is a function that describes the dependence on the angle $\vartheta$ between the particle's momentum $\mathbf{p}$ and the quantization axis (all Table \ref{tabla1}). One distinguishes different types of superfluidity, according to the angular momentum quantum numbers of the pairing wave function. Singlet $^{1}P_0$ pairs yield ``type A'' superfluidity, with an isotropic gap. The $^{3}P_2$ state yield ``type B'' and ``type C'' superfluidity. The description is rather uncertain because the energetically most favorable state of $n$-$n$ pairs ($|m_j|$ = 0, 1, or 2) is not known, being very sensitive to the $n$-$n$ pairs (\citealp{admun85}).
The critical temperature $T_c$ below which the matter becomes superfluid is related to the zero-temperature energy gap $\Delta(0)$ as given in Table \ref{tabla1}. 

It is useful to introduce the dimensionless temperature $\tau$ and dimensionless gap amplitude $v$: 
\begin{equation}
  \tau=\frac{T}{T_{c}}, \; v=\frac{\Delta(T)}{k_{B}T},
\end{equation}
where $k_{B}$ is the Boltzmann constant.

Moreover $v$ depends on the temperature by means of the BCS equation \citep{yak01}, whose solutions can be fitted by \citep{levyak94}:

\begin{eqnarray}
\label{eq:delta_a}
v_A= \sqrt{1-\tau}  \left[1.456-0.157\sqrt{ \frac{1}{\tau}} +1.764 \frac{1}{\tau}\right]\notag,
\end{eqnarray}
\begin{eqnarray}
\label{eq:delta_b}
v_B= \sqrt{1-\tau}  \left[0.7893-1.188 \frac{1}{\tau} \right]\notag,
\end{eqnarray}
\begin{eqnarray}
\label{eq:delta_c}
v_C=\frac{\sqrt{1-\tau^{4}}}{\tau}  \left[2.030-0.4903\tau^4+0.1727\tau^8\right] \notag .
\end{eqnarray}


\begin{table*}
\centering
\begin{tabular}{c c c c c c c} \hline
Model&$\Delta_{0} (Mev)$&$k_{1}(fm^-1)$&$k_{2}(fm^-2)$&$k_{3}(fm^-1)$&$k_{4}(fm^-2)$ & Reference\\ \hline
H & 4.8& 1.07 &1.8& 3.2 & 2 & \cite{modelHIJ} \\ 
I & 10.2& 1.09& 3& 3.45& 2.5& \cite{modelHIJ} \\ 
J &2.2& 1.05 &1 &2.82& 0.6 & \cite{modelHIJ} \\ 
K &0.425& 1.1& 0.5 & 2.7& 0.5 & \cite{modelKL} \\ 
L &0.068& 1.28& 0.1& 2.37 &0.02 & \cite{modelKL} \\ 
M &2.9& 1.21& 0.5& 1.62 &0.5 & \cite{modelM} \\ \hline
\end{tabular}
\caption[Superfluid $^{3}P_2$ models]{Parameters used in eq. \ref{eq:fenogap} for the gap models taken from \cite{anderson} and used in the present paper.}
\label{tabla2}
\end{table*}


Finally, in order to have a relation between the energy gap and the density (which we need to include the superfluid models in our code), we can also represent its amplitude (at the Fermi surface) by the phenomenological formula \citep{kam01}:



\begin{eqnarray}
 \label{eq:fenogap}
\Delta(k_F)=\Delta_0 \frac{(k_F-k_1)^{2}}{(k_F-k_1)^{2}+k_2}\frac{(k_F-k_3)^{2}}{(k_F-k_3)^{2}+k_4},
\end{eqnarray}
where $k_F$ is the Fermi momentum of the relevant nucleon,  and $k_1, k_2, k_3, k_4$ are parameters fitted for each theoretical energy gap model. Table \ref{tabla2} shows different models compiled by \cite{anderson} for $^{3}P_2$ pairing (superfluid type B or C).
Thus, we use the fact that
\begin{eqnarray}
 k_F=\frac{p_F}{\hbar}=(3\pi^{2} n)^{\frac{1}{3}},
\end{eqnarray}
where $n$ is the number density of the superfluid particle species.
\begin{figure}
\centering
\includegraphics[width=0.5\textwidth]{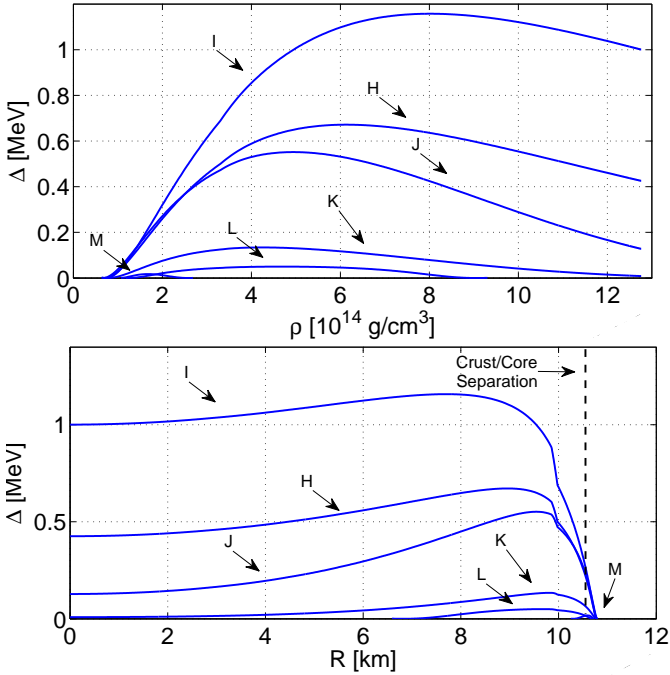}
\caption[]{Neutron superfluid energy gap as a function of density (upper panel) and of the radial coordinate (lower panel) in a neutron star of $1.76 M_{\odot}$ (as measured for PSR J0437-4715 by \citet{verbiest}, modeled with the $A18+\delta\nu+UIX^*$ EOS \citep{apr98}. The letters refer to various gap models compiled by \citet{anderson} and listed in Table \ref{tabla2}.}
\label{fig:models}
\end{figure} 

In Appendix \ref{sec:appendix1} we describe how we include this dependence in our code. In Figure \ref{fig:models},  we plot the energy gap as a function of density (upper panel) and as a function of the radial coordinate (lower panel) for all models. For a complete description of the models, see \cite{anderson} and references therein.

\subsection{Neutrino emissivity}
\label{neutrino-section}

Once allowed by momentum conservation, the direct Urca (Durca) reactions are the dominant neutrino emission processes. Indeed, \citet{petro11} showed that when including Durca reactions in models of rotochemical heating the evolution changes dramatically: no quasi-steady state is reached and the star spends most of its lifetime at low surface temperatures ($<10^4$ K). Whether Durca processes are allowed or not depends on the EOS and on the stellar mass. For simplicity, we ignore their effect and follow PR10 in considering only the modified Urca (Murca) reactions \citep{yak01}:
\begin{eqnarray}
\label{eq:nbranch}
n+n_i\rightarrow p+n_f+l+\bar{\nu_{l}},\qquad p+n_i+l\rightarrow n+n_f+\nu_{l};\\
\label{eq:pbranch}
n+p_i\rightarrow p+p_f+l+\bar{\nu_{l}},\qquad p+p_i+l\rightarrow n+p_f+\nu_{l};
\end{eqnarray}
where the subindices $i$ and $f$ represent the initial and final states of the particle, and $l=e,\mu$. Equations (\ref{eq:nbranch}) and (\ref{eq:pbranch}) coresspond to the so-called neutron and proton branch, respectively.
We write the neutrino luminosity and the net reaction rate due to Murca reactions involving the lepton $l$ and integrated over 
the core, respectively, as
\begin{eqnarray}
 L_{\nu,l}^{\infty}= \tilde L_{nl}I^{n}_{M,\epsilon}T^{8}_{\infty} + \tilde L_{pl}I^{p}_{M,\epsilon}T^{8}_{\infty},\\
 \Delta\tilde\Gamma_{npl}^{\infty}= \frac{\tilde L_{nl}}{k_{B}} I^{n}_{M,\Gamma}T^{7}_{\infty} + \frac{\tilde L_{pl}}{k_{B}}I^{p}_{M,\Gamma}T^{7}_{\infty},
\end{eqnarray}
where the upper $n$ and $p$ stand for the two Murca branches, and the constants $\tilde L_{nl}$ and $\tilde L_{pl}$ are defined in terms of the
neutrino luminosities for a non-superfluid NS in beta equilibrium (for details see PR10).
The quantities $I^{N}_{M.\epsilon}$ and $I^{N}_{M.\Gamma}$ are dimensionless phase-space integrals that contain the dependence
of the emissivity and the net reaction rate, respectively, on the chemical imbalances $\eta^{\infty}_{npl}$ and on the energy gaps.
To introduce these integrals, it is useful to define the usual dimensionless variables normalized by the thermal energy $k_{B}T$, as
\begin{eqnarray}
 x_j\equiv\frac{\epsilon_j-\mu_j}{k_{B}T},\;x_\nu\equiv\frac{\epsilon_\nu}{k_{B}T},\; \mbox{and} \;\xi_l\equiv\frac{\eta_{npl}}{k_{B}T},
\end{eqnarray}
which represent the energy of the non-superfluid degenerate particle $j$, the neutrino, and the chemical imbalance involving the lepton $l$, respectively, while for the superfluid nucleon $i$ we write:
\begin{eqnarray}
x_i\equiv\frac{v_{F_i}(p_i-p_{F_i})}{k_{B}T},\; \mbox{and} \; z_i\equiv \sgn(x_i)\sqrt{x_i^{2}+\delta_i^{2}}
\end{eqnarray}
where $v_{F_i}$ and $p_{F_i}$ are the Fermi velocity and the Fermi momentum, respectively.
In terms of these variables, for the case of the neutron branch Murca reactions
and considering the most general case (neutrons and protons as superfluids), the integrals are:
\begin{eqnarray}
\label{eq:murca_inicial}
I^{N}_{M,\Gamma}&=&\frac{1}{(4\pi)^5}\int d\Omega_{n}d\Omega_{n_{i}}d\Omega_{n_{f}}d\Omega_{p}d\Omega_{e} \int_0^{\infty}dx_{\nu} x_{\nu}^2 \\ \notag
&&\times \int_{-\infty}^{\infty}dx_{n}dx_{n_i}dx_{n_f}dx_{p}dx_{e}\\ \nonumber
&&\times f(z_n)f(z_{n_i})f(z_{n_f})f(z_p)f(x_e) \\ \nonumber
&&\times  \{\delta(x_{\nu}+\xi_l-z_n-z_{n_i}-z_{n_f}-z_p-x_e)  \\ \nonumber
&&- \delta(x_{\nu}-\xi_l-z_n-z_{n_i}-z_{n_f}-z_p-x_e) \} \\
\label{eq:murca_inicial2}
I^{N}_{M,\epsilon}&=&\frac{1}{(4\pi)^5}\int d\Omega_{n}d\Omega_{n_{i}}d\Omega_{n_{f}}d\Omega_{p}d\Omega_{e}\int_0^{\infty}dx_{\nu} x_{\nu}^3 \\ \notag
&&\times\int_{-\infty}^{\infty} dx_{n}dx_{n_i}dx_{n_f}dx_{p}dx_{e} \\ \nonumber
&&\times f(z_n)f(z_{n_i})f(z_{n_f}) f(z_p)f(x_e)\\ \nonumber
&&\times \{\delta(x_{\nu}+\xi_l-z_n-z_{n_i}-z_{n_f}-z_p-x_e)\\ \nonumber
&&+ \delta(x_{\nu}-\xi_l-z_n-z_{n_i}-z_{n_f}-z_p-x_e) \},
\end{eqnarray}
where $f(x)=1/(1+e^{x})$ is the Fermi function and $d\Omega_{k}$ is the solid angle element in direction of $\mathbf{p}_{k}$: 
\begin{eqnarray}
\int \frac{d\Omega_{k}}{4\pi}=\int_0^{\pi/2} \sin(\vartheta_k) d\vartheta_k,
\end{eqnarray}
with $k$ the index of the particles involved, $k=n,n_i,n_f,p,e$. 

In the non-superfluid case (i.e., $\delta_n=\delta_p=0$), these integrals reduce to the polynomials calculated by \cite{reis95}:
\begin{eqnarray}
 I^{N}_{M,\epsilon} (\delta_n=\delta_p=0)\equiv F_M(\xi_l)= \nonumber
\end{eqnarray}
\begin{eqnarray}
 1 + \frac{22020\xi_{l}^2}{11513\pi^2} + \frac{5670\xi_{l}^4}{11513\pi^4} + \frac{420\xi_{l}^6}{11513\pi^6} + \frac{9\xi_{l}^8}{11513\pi^8},
\end{eqnarray}
\begin{eqnarray}
 I^{N}_{M,\Gamma} (\delta_n=\delta_p=0))\equiv H_M(\xi_{l})= \nonumber
\end{eqnarray}
\begin{eqnarray}
 \frac{14680\xi_{l}}{11513\pi^2}+\frac{7560\xi_{l}^3}{11513\pi^4}+\frac{840\xi_{l}^5}{11513\pi^6} +\frac{24\xi_{l}^7}{11513\pi^6}.
\end{eqnarray}

The Cooper pairing reduces the emissivities and net reaction rates. A way to account for this, is to define the so-called reduction factors as the ratio of these superfluid integrals and their non-superfluid limits:
\begin{eqnarray}
R^{N}_{M,\epsilon}(\xi_l,\delta_n,\delta_p)&=&\frac{I^{N}_{M,\epsilon}(\xi_l,\delta_n,\delta_p)}{F_M(\xi_l)} 
\label{eq:reduction}
\end{eqnarray}
\begin{eqnarray}
R^{N}_{M,\Gamma}(\xi_l,\delta_n,\delta_p)&=&\frac{I^{N}_{M,\Gamma}(\xi_l,\delta_n,\delta_p)}{H_M(\xi_l)}.
\label{eq:reduction2}
\end{eqnarray}
The most time-consuming computation in the evolution of the star are these factors. In order to calculate them we have to analyse different evolutionary stages because of their strong dependence on the chemical imbalances, temperature, and superfluid gaps. In Appendix  \ref{sec:appendix_1} we explain the numerical approach that we have used to manage these calculations. 
\subsection{Pair breaking \& pair formation emission}
\begin{figure*}
\includegraphics[width=0.82\textwidth]{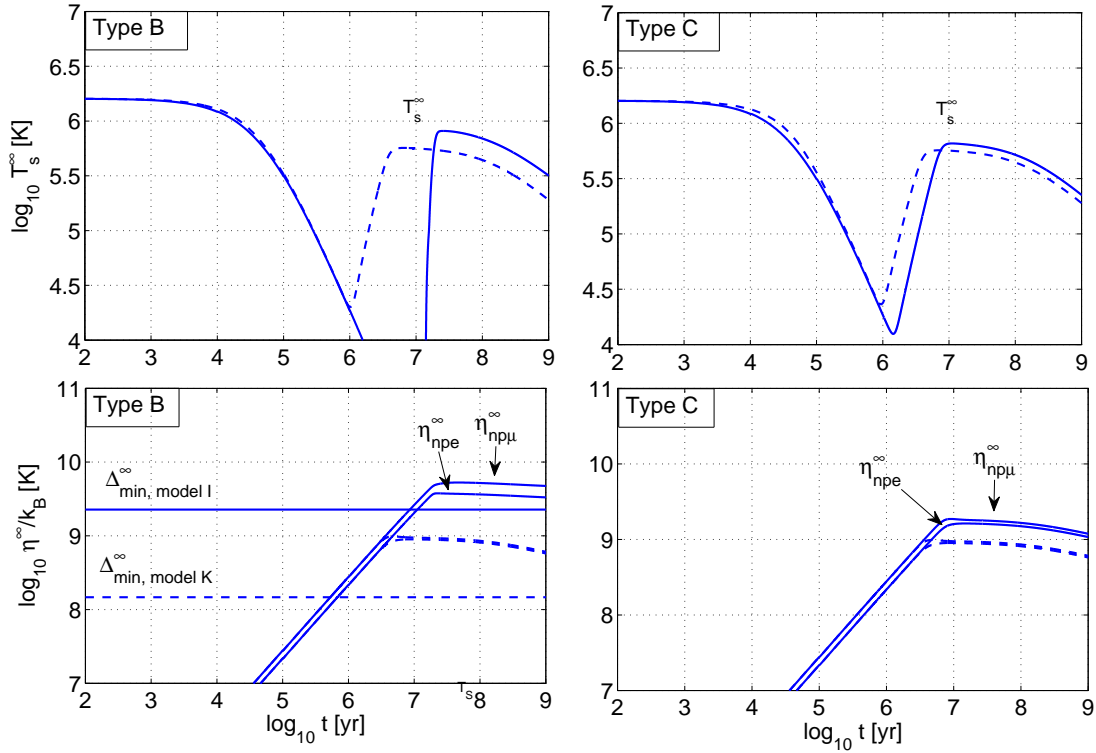}
\caption[]{Thermal evolution of a $1.76 \mathrm{M_{\odot}}$ MSP with magnetic field  $B=3.275 \times10^{8} \;\mathrm{G}$ and initial period $P_0=1\;\mathrm{ms}$ modeled with the $A18+\delta\nu+UIX^*$ EOS \citep{apr98}. 
Left and right panels: Type B and C superfluidity, respectively. Upper panels: Thermal evolution for model I (solid lines) and K (dashed lines). Lower panels: Chemical evolution of both chemical imbalances ($\eta_{np\mu}$ and $ \eta_{npe}$) for the same superfluid model. The horizontal lines in the lower left panel are the minimum gap of the respective superfluid model (see Fig. \ref{fig:models}). All the curves were calculated using the light elements envelope model.}
\label{fig:modelos_desbalance}
\end{figure*} 
\label{sec:pb}
When the temperature is just below the superfluid transition temperature $T_c$, new channels for neutrino emission become operative, namely the pair-breaking and pair-formation (PBF) processes first proposed by \cite{flowers}. Therefore, they are important in young neutron stars with high initial temperatures, e.g., classical pulsars (CPs), but not in old, cool objects such as millisecond pulsars (see also PR10). 
We take the emissivities for these PBF processes as \citep{yak01}:
\begin{eqnarray}
Q_{PBF}= \nonumber
\end{eqnarray}
\begin{eqnarray}
1.46 \times 10^{22} \left( \frac{m^{*}_{N}}{m_N}\right)  \left( \frac{p_{F} }{m_N c}\right) \; T_{9}^{7} \; F(\delta) \; \mathrm{erg \; cm^{-3} \; s^{-1} }
\end{eqnarray}
with $m^{*}_{N}$ the effective mass of the nucleon $N$, which is a function of density, and the function $F(\delta)$ given by:
\begin{eqnarray}
F_{A}(\delta_{A}) &=& (0.602 \; \delta_{A}^2  + 0.5942 \;\delta_{A}^4 + 0.288 \; \delta_{A}^6)   \nonumber \\ &&\times \left( 0.5547 + \sqrt{(0.4453)^{2} + 0.0113 \;\delta_{A}^{1/2}}\right)^{2}\nonumber 
\\ &&\times \exp\left( -\sqrt{4 \delta_{A}^{2} + (2.245)^{2}} + 2.245 \right),
\end{eqnarray}
\begin{eqnarray}
F_{B}(\delta_{B}) &=& \frac{ 1.204 \; \delta_{B}^{2} + 3.733 \; \delta_{B}^{4} + 0.3191 \;\delta_{B}^{6} }{1+ 0.3511 \; \delta_{B}^{2}}  \nonumber \\ &&\times \left( 0.7591 + \sqrt{(0.2409)^{2} + 0.3145 \delta_{B}^{2}}\right)^{2}\nonumber 
\\ &&\times \exp\left( -\sqrt{4 \delta_{B}^{2} + (0.4616)^{2}} + 0.4616 \right),
\end{eqnarray}
\begin{eqnarray}
F_{C}(\delta_{C}) &=&(0.4013 \; \delta_{C}^{2} + 0.043\; \delta_{C}^{4}+ 0.002172\;\delta_{C}^{6}) \\ \notag
&&(1-2.018\times 10^{-1} \; \delta_{C}^{2} + 2.601\times 10^{-2}\; \delta_{c}^{4}\\ \nonumber
&&-1.477 \times 10^{-3}  + 4.34\times 10^{-5}\;\delta_{c}^{8} )^{-1},
\end{eqnarray}
for the three types of superfluidity discussed in \S\ref{sec:super}.
\subsection{Specific heat}
When $T$ decreases, crossing $T_c$, there is a discontinuous increase in the specific heat, characteristic of a second-order phase transition. When $T\ll T_c$, an exponential-like suppression occurs due to the presence of the gap in the energy spectrum. These effects are taken into account by using control functions ($C_{f}$) that multiply the unpaired values of the specific heat at constant volume $C_V$ (see PR10 for details).

We use a fit made by \cite{levyak94}, as given by \cite{yak99}:
\begin{eqnarray}
  C_{f_{A}}= \left[0.4186+\sqrt{(1.007)^2+(0.5010\; \delta_{B})^2}\right]^{2.5}\times \nonumber \\
\times \; \exp(1.456-\sqrt{1.456^2+\delta_{A}^2});
\end{eqnarray}
\begin{eqnarray}
  C_{f_{B}}= \left[0.6893+\sqrt{0.790 ^2+(0.2824\; \delta_{B})^2}\right]^2\times \nonumber \\
\times \; \exp(1.934-\sqrt{1.934^2+\delta_{B}^2});
\end{eqnarray}
\begin{eqnarray}
 C_{f_{C}}= \frac{2.188-(9.537 \times 10^{-5} \;\delta_{C})^2 + (0.1491\;\delta_{C})^4 }{ 1+(0.2846\;\delta_{C})^2 + (0.01335\;\delta_{C})^4 + (0.1815 \delta_{C})^6 }.
\end{eqnarray}

\section{Results and Discussion}
\label{sec:results}

Since we are considering two different classes of pulsars, with very different spin-down histories, we will study them separately in \S3.1 and \S3.2.

\subsection{Millisecond pulsar regime}

In this section, we study the effects of rotochemical heating in MSPs. These objects have inferred dipole magnetic fields $\sim10^{8-9} \;\mathrm{G}$, initial period (after $recycling$) $\sim1-5\;\mathrm{ms}$, and ages $\sim10^{8-10}\; \mathrm{yr}$. We compare our calculations with the PSR J0437-4715, whose measured mass ($[1.76\pm 0.20]\;\mathrm{M_{\odot}}$; \citet{verbiest}) we also use as our reference value.


One of the predictions of rotochemical heating with superfluidity is that the neutrino reactions that heat the star will be suppressed until the chemical imbalances overcome a threshold imposed by the minimum gap \citep{reis97}. At this point, reactions will occur easily, and each will release an amount of energy equivalent to the chemical imbalance. Thus, the larger the gap, the higher the temperature obtained.

Following this argument, and observing the form of $F(\vartheta)=1+\cos^{2}\vartheta$ and Equation (\ref{eq:deltagap}), we expect that for type B superfluidity the temperature will be higher than for type C superfluidity, which has $F(\vartheta)=\sin^{2}\vartheta$, because the gap amplitude for type B superfluidity has a higher value for every angle in $F(\vartheta)$ (compared with type C), which implies a higher threshold. 

In order to estimate the surface temperature from the internal temperature, it is necessary to use a model for the outermost layers of the neutron star, where most of the temperature drop occurs. In the literature there are two models that are often used: the \cite{gpe83} envelope model, which assumes heavy (iron-like) elements in the atmosphere and surface of the star, and the \cite{potek} envelope model, which instead uses light elements, primarily H and He. Since we are interested in the late stage of the evolution of this MSP, changing the envelope will not change our results since in the quasi-steady state all the energy dissipated in the star is released through the surface. What governs the generation of energy are the chemical reactions that are driven by the chemical imbalances, and because the chemical imbalance is determined by the spin-down and not by the chemical composition of the envelope, the star has the same surface temperature using both envelopes. In this regime, we choose the light-
element 
model in our plots.
\begin{figure}
\includegraphics[width=0.5\textwidth]{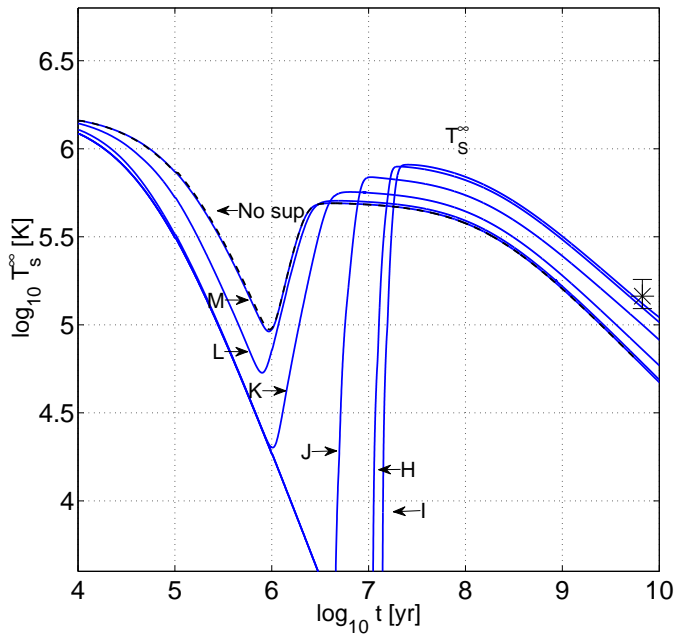}
\caption[]{Thermal evolution of a  $1.76 \mathrm{M_{\odot}}$ MSP with magnetic field  $B=3.275 \times10^{8} \;\mathrm{G}$ and initial period $P_0=1\;\mathrm{ms}$ modeled with the $A18+\delta\nu+UIX^*$ EOS \citep{apr98}. The solid curves correspond to the six neutron gap models, using type B superfluidity taken from \cite{anderson}. The dashed one is for a non-superfluid NS. All the curves were calculated using the light elements envelope model from \cite{potek}. The point with error bar is the measurement of PSR J0437-4715 from \cite{Durant}, assuming a black body fit, and with the error bar reflecting mostly the assumed range of radii (see \S\ref{sec:conclusions} for a discussion).} 
\label{fig:modelos_msp_tipoB}
\end{figure} 
\begin{figure}
\includegraphics[width=0.5\textwidth]{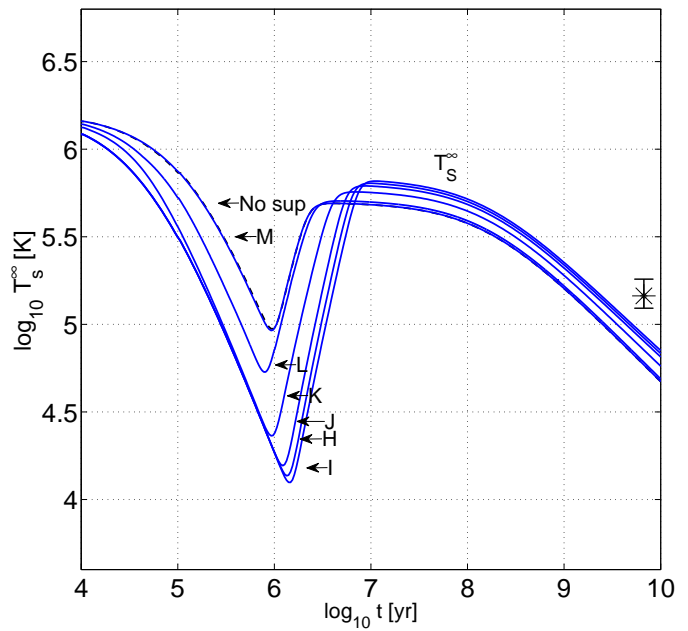}
\caption[]{The same as Fig. 4, but for superfluidity type C.}
\label{fig:modelos_msp_tipoC}
\end{figure}

To explore the effects of anisotropic superfluidity in this heating mechanism, we plot in Figure \ref{fig:modelos_desbalance} the evolution of a MSP with type B and C superfluidity for two superfluid models, I (with a large gap, see Figure \ref{fig:models}) and K (with a much smaller gap). By observing the left and right upper panel of the same superfluid model (particulary model I), it is clear that type B superfluidity predicts a higher temperature than type C. Also, for a given type of superfluidity, the temperature obtained will increase with the superfluid gap.

The cause of this is hidden in the chemical imbalances. For type B superfluidity, after the chemical imbalances overcome the minimum gap threshold $\Delta_{min}$, the chemical reactions are allowed, causing an increase in the internal temperature of the star. In the case of type C, the reactions are always allowed since the gap vanishes in one direction, so it is not necessary for the chemical imbalances to overcome a threshold. Therefore, there is no complete blocking of the chemical reactions as in type B, so the heating stage will be reached when the chemical reactions that heat the star overcome the cooling due to photons and neutrinos. In both cases, the star reaches a quasi-steady state, as expected.

Figures \ref{fig:modelos_msp_tipoB} and \ref{fig:modelos_msp_tipoC} show our results for the surface temperature for all six models, calculated with superfluity type B and C respectively. Only the models with the largest gaps (I and H) and with type B superfluidity can explain the surface temperature of the PSR J0437-4715. This conclusion is valid independently of the EOS\footnote{We tested this using the same EOSs as PR10: A18+$\delta v$ +UIX, A18+$\delta v$, BPAL11, BPAL21, BPAL31, and BPAL33.}, and also of the uncertainty in the MSP's mass.

\subsection{Classical pulsar regime}

The previous analysis raises the question: Given that the two models with the largest gaps and type B superfluidity can explain the surface temperature of J0437-4715, can these same models also explain the available observations of classical pulsars?

To answer this question, we need to consider different aspects that were neglected in the MSP regime. The first one is the pair breaking and formation process explained in section \S\ref{sec:pb}. The net effect of this process is an enhancement of the cooling at an early stage, when the temperature has just dropped below the transition temperature. Another aspect to take into account is the envelope model. In this regime we analyse both models proposed in the literature, which do cause important changes in the predicted temperature in the CP regime, as we will show below. Additional factors that can affect the shape of the curves with rotochemical heating are the mass, the magnetic field, and the initial period of the star. The sample of observations was taken from \cite{yak08}, who collect three upper limits and thirteen surface temperature measurements of young neutron stars. We also add the 
upper limits for the older objects B1929+10, B1133+16, B0950+08, J0108-143 and J2144-3933 listed by \cite{yak01}. 
\begin{figure}
\includegraphics[width=0.5\textwidth]{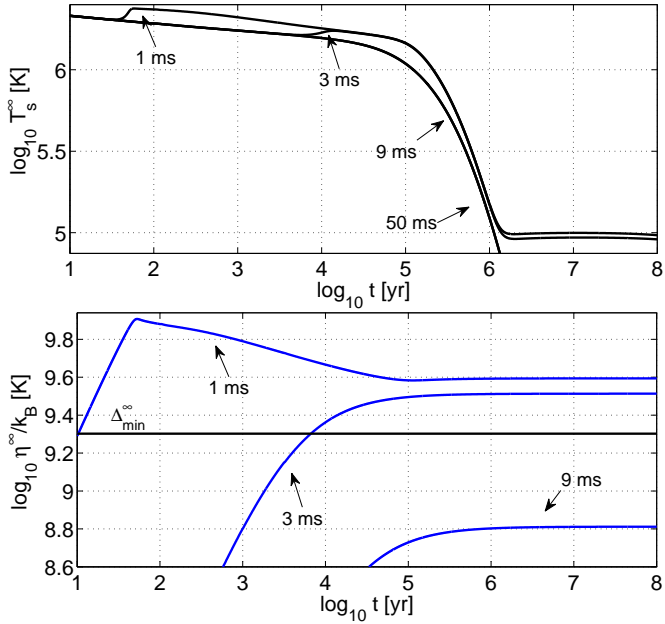}
\caption[]{Thermal evolution of a neutron star of 1.4 $M_\odot$, $B=2.5\times 10^{11}\;\mathrm{G}$ with model H superfluidity type B and heavy elements envelope model modeled with the $A18+\delta\nu+UIX^*$ EOS \citep{apr98}. Upper panel: Evolution of the object with different initial periods $P_{0}=1, 3, 9, 50 \;\mathrm{ms}$ (the latter two are indistinguishable on the plot). Lower panel: Chemical imbalance $\eta_{np\mu}$, for three thermal evolutions, with $P_{0}=1,3,9 \;\mathrm{ms}$ respectively. The horizontal black line is the minimum gap of the model.} 
\label{fig:periodo_y_B_conbox}
\end{figure} 
\begin{figure}
\includegraphics[width=0.5\textwidth]{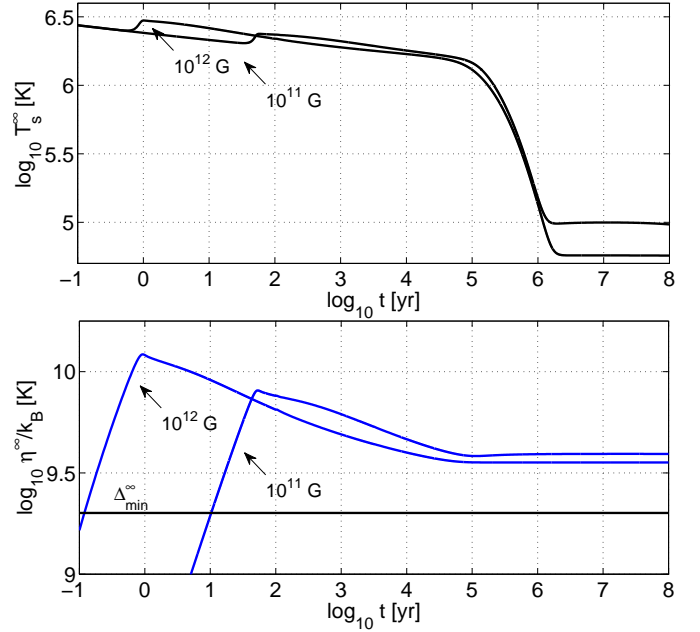}
\caption[]{Thermal evolution of a neutron star of 1.4 $M_\odot$, and $P_0=1\;\mathrm{ms}$, with model H, superfluidity type B and heavy elements envelope model modeled with the $A18+\delta\nu+UIX^*$ EOS \citep{apr98}. Upper panel: Evolution of the object with different magnetic fields of  $B= 10^{11}, 10^{12}\;\mathrm{G}$. Lower panel: Chemical imbalance $\eta_{np\mu}$, of the same objects. The horizontal black line is the minimum gap of the model.} 
\label{fig:B_conbox}
\end{figure}

\begin{figure}
\includegraphics[width=0.5\textwidth]{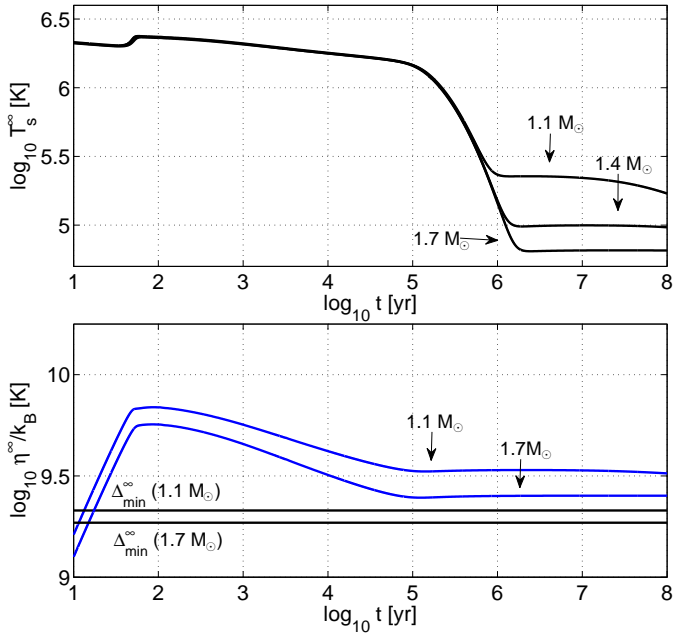}
\caption[]{Thermal evolution of a neutron star with an initial period of $P_0=1\;\mathrm{ms}$ a magnetic field of $B=2.5 \times 10^{11}\;\mathrm{G}$, superfluidity type B and the superfluid model H modeled with the $A18+\delta\nu+UIX^*$ EOS \citep{apr98} with type B superfluidity. Upper panel: Different mass values $M= 1.1, 1.4, 1.7\; M_{\odot}$, with superfluidity type B. Lower panel: Chemical imbalance $\eta_{np\mu}$ of two objects with $M= 1.1$ and, $1.7\;M_{\odot}$ respectively. Also is plotted the minimum gap of the model. The fact that there are two minimum gaps is due to the effect of gravitational redshift which depends on the mass of the star.}
\label{fig:masa_doble}
\end{figure}

It is important to note that the effective dipole magnetic field is known for all of these stars, whereas their mass and initial period are unknown variables. We also have the current period of the objects, and thus their age for a given initial period and spin-down model (which we always take as a pure dipole, with braking index $n=3$). Therefore, in order to know if a certain superfluid model can explain all the observations, we need to explore the behavior when  changing each of the relevant parameters. We start by studying the models with effectively the highest energy gaps (I and H). For the sake of clarity we restrict our analyisis in Figs. \ref{fig:periodo_y_B_conbox}, \ref{fig:B_conbox}, and \ref{fig:masa_doble} to the model H with type B superfluidity, one chemical imbalance ($\eta_{np\mu}$), and an envelope with heavy elements. The effects of changing the gap and envelope models will be explained afterwards.


We begin by varying the initial period (Figure \ref{fig:periodo_y_B_conbox}). As we lengthen it (upper panel), the effect of rotochemical heating on the surface temperature is delayed, therefore, for a longer initial period the temperature at a given age is generally lower. 
Clearly, for the shorter period, the chemical imbalances (lower panel) grow fast and at an early time, so they soon become larger than the energy gap, at which point strong Murca reactions are turned on, stopping the growth and generating heat inside the star. For the case of type B superfluidity shown in the figure, the curves with long initial period never exceed the threshold, causing the reactions that heat the star to be severely suppressed. For type C superfluidity (not shown), the reactions are more mildly suppressed causing a similar but weaker effect. 

The effect of varying the magnetic field (Figure \ref{fig:B_conbox}) is important in the early and late stages of the evolution. For a strong magnetic field, the chemical imbalances grow very fast at the beginning, so they overcome the threshold quickly, causing the same effect as with short initial periods: Murca reactions are turned on, stopping the growth and generating heat inside the star. On the other hand, a strong magnetic field makes the star spin down quickly, so at late times there will be much less rotochemical heating than for lower magnetic fields. This effect is not modified by the superfluidity and the envelope model. 

The effect of increasing the mass (Figure \ref{fig:masa_doble}) is noticeable in the late stages of the evolution ($t>10^6 \mathrm{yr}$) for type B superfluidity (and in the early stages for type C). In both cases a higher mass results in a lower surface temperature. 

Based on these results, we conclude that the highest temperatures are predicted for the shortest initial periods, the lowest magnetic fields, and the lowest masses, while the lowest predicted temperatures follow the opposit trend. Thus, we generate a high-temperature prediction choosing a very short initial period, $P_0=1\;\mathrm{ms}$, the lowest magnetic field in our sample, $B=2.51 \times 10^{11} \;\mathrm{G}$, and the lowest neutron star mass observed so far, $1.25\; M_{\odot}$ \citep{lyne}. Similarly, a low-temperature prediction is generated with the longest initial period, which we choose as $P_{0}=50\;\mathrm{ms}$, the highest magnetic field of our sample, $B=9.3 \times 10^{12}\;\mathrm{G}$, and a mass of $1.97 \;\mathrm{M_{\odot}}$, consistent with the highest that have been measured precisely until now, $(1.97\pm 0.04)\;\mathrm{M_{\odot}}$ for PSR J1614-2230 \citep{demo} and $(2.01\pm 0.04)\;\mathrm{M_{\odot}}$ for  PSR J0348+0432 \citep{anto13}. All allowed combinations of parameters 
should yield predictions lying between these two curves. Of course, for any specific object, the magnetic field should be taken as the measured one, which would reduce 
the  
temperature range acceptable for it.

\begin{figure*}
\includegraphics[width=0.8\textwidth]{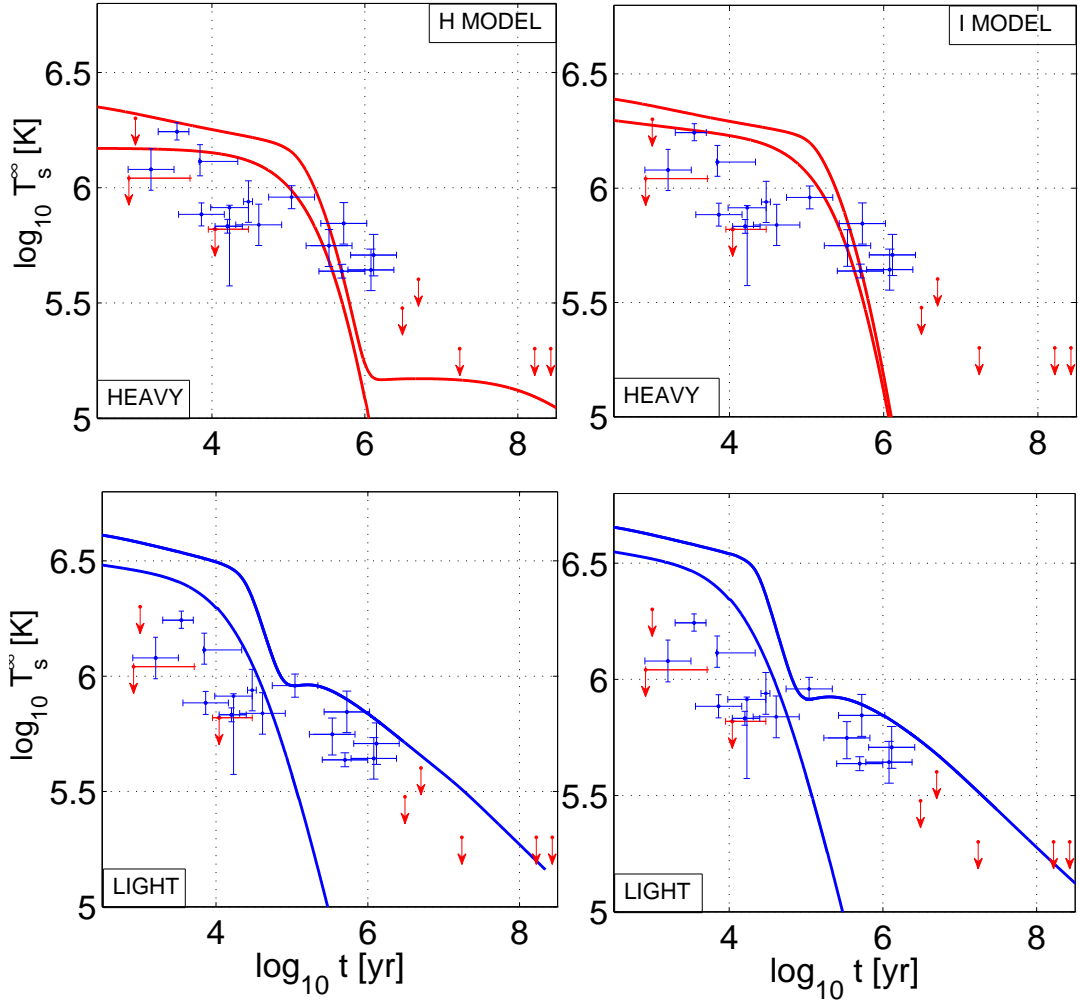}
\caption[]{Upper and lower limits of the predicted surface temperature calculated using the $A18+\delta\nu+UIX^*$ EOS \citep{apr98} and uperfluidity type B for all models. The two upper panels (red lines) use the heavy elements envelope model, while the two lower panels were calculated using the light elements envelope model. Everything outside the limits, cannot be explained with the models. All the observations plotted are classical pulsar taken from \cite{yak08}}
\label{fig:cotas_tipoB}
\end{figure*}

\begin{figure*}
\includegraphics[width=0.8\textwidth]{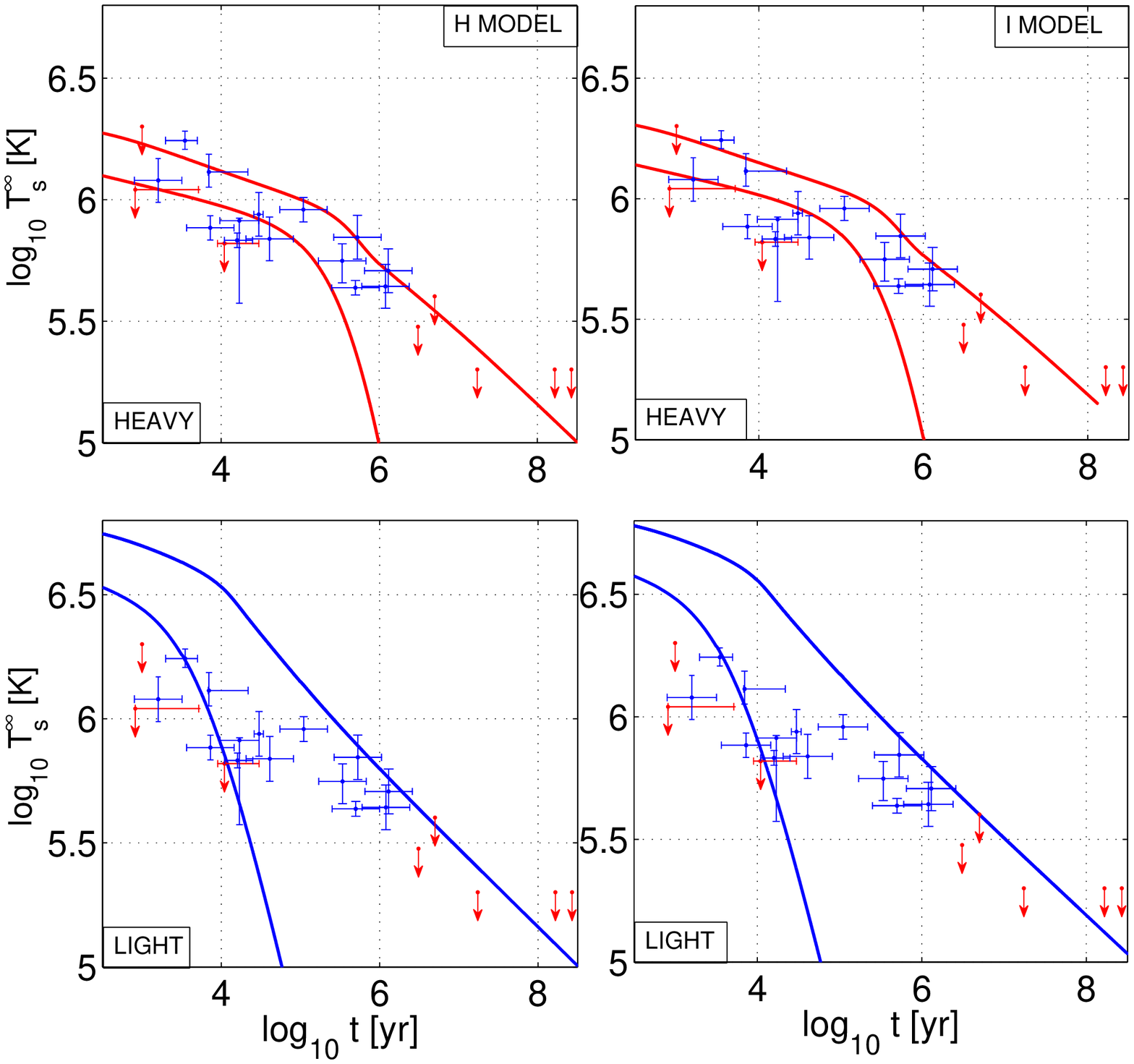}
\caption[]{Upper and lower limits of the predicted surface temperature calculated using the $A18+\delta\nu+UIX^*$ EOS \citep{apr98} and uperfluidity type C for all models. The two upper panels (red lines) use the heavy elements envelope model, while the two lower panels were calculated using the light elements envelope model. Everything outside the limits, cannot be explained with the models. Using both types of envelope the two models are able to explain the observations}
\label{fig:cotas_tipoC}
\end{figure*}

These curves are shown in Figs. \ref{fig:cotas_tipoB} and \ref{fig:cotas_tipoC}, for different combinations of gap model (I or H, both with large gaps), envelope composition (light or heavy elements), and type of superfluidity (B in Fig. \ref{fig:cotas_tipoB}, C in Fig. \ref{fig:cotas_tipoC}). The available observational data (measurements and upper limits) are also shown on each panel. Clearly, for type B superfluidity (Figure \ref{fig:cotas_tipoB}), all combinations overpredicts the temperature of many of the young neutron stars. For type C superfluidity (Figure \ref{fig:cotas_tipoC}), all objects can in principle be fitted, as long as we allow for different envelope compositions in different objects.
\newpage
\newpage
 \newpage
\section{Conclusions}\label{conclusions}
\label{sec:conclusions}
We have studied the rotochemical heating effect in two regimes, for classical and millisecond pulsars with modified Urca reactions in the presence of density-dependent anisotropic Cooper pairing gaps for the neutrons (but no gaps for the protons). We calculate the surface temperature in those regimes using the two anisotropic types of superfluidity, type B and C, considering six superfluid models taken from \cite{anderson}, also allowing for two different envelope models proposed in the literature, one composed of heavy elements \citep{gpe83}, and the other of lighter elements \citep{potek}.

In the millisecond pulsar regime, we fit the temperature of the PSR J0437-4715, recently measured more precisely by \cite{Durant}. This object provides us a constraint on the surface temperature calculated with the models. Among the six superfluid models studied, only the two with the largest gaps, I and H, and type B superfluidity predict a high enough temperature to agree with the allowed range reported by  \cite{Durant}, while none of the models does so for type C superfluidity.
 
For the classical pulsar regime, we take the two models with the largest gaps, I and H, and constrain them using twenty-one observations, eight of which are upper limits. We discard the type B superfluidity as the most probable way to explain the observations. Type C superfluidity, instead, might be able to explain all the objects, if either envelope model can be chosen arbitrarily for each object. 

Thus, there does not appear to be a single model for neutron superfluidity that can simultaneously explain all available observations of the thermal emission of non-accreting neutron stars (in particular, those of the PSR J0437-4715 \emph{and} all available classical pulsars). There are several possible explanations for this inconsistency:

1) The temperature range given by \cite{Durant} for the PSR J0437-4715 is based on a black body fit to the ultraviolet (UV) emission for a spherical object with ``radiation radius'' $R_\infty$ between two values: the smallest radius that does not overpredict the X-ray emission of this object (which might be largely non-thermal),  $R_\infty=7.8\;\mathrm{km}$ and a ``fiducial radius''  $R_\infty = 15\;\mathrm{km}$. There are a few potential or actual problems with this (\citealt{denix2}). On the one hand, the actual spectral energy distribution might differ from a black body, thus yielding a different temperature for a given UV flux. In fact, the temperature inferred for an iron atmosphere is essentially the same as for a black body, whereas a helium atmosphere yields a somewhat higher temperature (by a factor of 1.3, for the same assumed $R_\infty$), and a hydrogen atmosphere a much higher temperature (factor of 2.6), which however overpredicts the X-ray emission for expected 
neutron star radii $R_\infty<20\;\mathrm{km}$. On the other hand, the lower limit $R_\infty=7.8\;\mathrm{km}$ (yielding the upper limit for the temperature), to be consistent with General Relativity, would require a stellar mass $<1.2 \;M_\odot$, far below the reported measurement ($1.76 \pm 0.20\;M_\odot$; \citealt{verbiest}), and quite implausible for MSPs, which are believed to be ``recycled`` through accretion. Moreover, the upper limit $R_\infty=15\;\mathrm{km}$ (responsible for the lower limit for the temperature) might be too conservative. In fact, \citet{heb13} find that the combination of the best available theoretical and observational constraints (the latter from the recent mass measurements $\approx 2 \;M_\odot$) allows NS coordinate radii up to $R\sim 14$ km, corresponding to radiation radii $R_\infty\sim 18$ km, and \citet{bog13}, based on X-ray spectra and light curves of PSR J0437-4715, assumed to have a hydrogen atmosphere at its hot polar caps, obtains 
a 3 $\sigma$ lower bound $R>11.1$ km, corresponding to $R_\infty \gtrsim 15.3$ km. Thus, the actual temperature of this MSP might be somewhat lower than the range given by \cite{Durant}, perhaps allowing for superfluid models with smaller gaps or type C superfluidity, although formally the change is insufficient.

2) Some of the young neutron stars with relatively low temperatures might in fact have fast cooling processes (Durca or driven by ``exotic'' particles) in their inner core, which are somewhat suppressed by superfluidity \citep{yak04}. However, the ``recycling'' scenario would imply that MSPs, having undergone substantial accretion, should generally have higher masses, and thus be more likely subject to these processes than classical pulsars. 

3) In order to simplify the numerical problem, in our calculations we ignored the possibility of having proton superconductivity coexisting with neutron superfluidity, contrary to what is invoked (\citealp{Shternin} and \citealp{page11}) to explain the apparent fast cooling observed in the hot neutron star in the supernova remnant Cas A (\citealp{heinke}; \citealp{elsha}). A large proton gap would suppress Urca reactions and keep the star as hot as it is to its present age, but a moderate neutron gap would be responsible for pair breaking and formation processes accounting for the present, fast cooling. We note, however, that the claimed observation of the fast cooling was questioned by \citealp{posselt}, which would invalidate the argument.

Rotochemical heating is unlikely to be important for the very young neutron star in Cas A, but it is undoubtedly important for older classical pulsars, and particularly for MSPs such as PSR J0437-4715. Thus, an obvious, but challenging next step is to study the thermal evolution of neutron stars considering both the superfluid-superconductor models consistent with Cas A and the effects of rotochemical heating. 

\section*{Acknowledgements}
We thank Denis Gonz\'alez for useful conversations and advice, Dima Yakovlev for clarifications about his work, and an anonymous referee for detailed and useful comments that improved the manuscript. This work was supported by FONDECYT Regular Projects 1060644 and 1110213, Proyecto Basal PFB-06/2007, and CONICYT International Collaboration Grant DFG-06.


\clearpage
\appendix
\onecolumn

\section{Modeling strategy}
\label{sec:appendix1}
\label{sec:density}

In order to calculate the time-evolution of the temperature of a neutron star, the first step is to evaluate the reduction factors (Equations (\ref{eq:reduction})-(\ref{eq:reduction2})) considering anisotropic superfluidity, and later include them in a density dependence scheme. To do so, a five-dimensional integral needs to be computed, and only
one dimension can be eliminated by integrating out analytically the electron variable in Equations \ref{eq:murca_inicial} and \ref{eq:murca_inicial2}. However, if one of the nucleons is not superfluid, as we will assume below, it is possible to eliminate more dimensions by integrating out, analytically as well, the non-superfluid 
variables. In addition, neutron and proton branches (Equations (\ref{eq:nbranch}) and (\ref{eq:pbranch})) are nearly equal in the absence of superfluidity or when the energy gaps are small (or similar to each other). However, when one particle has a substantially smaller (or zero) gap, the corresponding branch will be strongly dominant. If we only consider the neutron as a superfluid particle, the neutron branch is strongly suppressed and can be neglected to first approximation. For the proton branch we can integrate out the electron and the three non-superfluid protons, obtaining a two-dimensional integral that has to be calculated numerically (see Appendix \ref{sec:appendix2}).

To incorporate the density-dependence of the energy gap in the formalism developed by PR10 with uniform energy gaps we proceed as follows: (i) we divide the core in twelve or less regions in which the amplitude of the gap have substantial variation, (ii) we take an average of the amplitude within each shell, (iii) we calculate the luminosities, specific heat, and reduction factors.  
\section{Gap anisotropy }
\label{sec:appendix2}
\begin{figure*}
\includegraphics[width=0.65\textwidth]{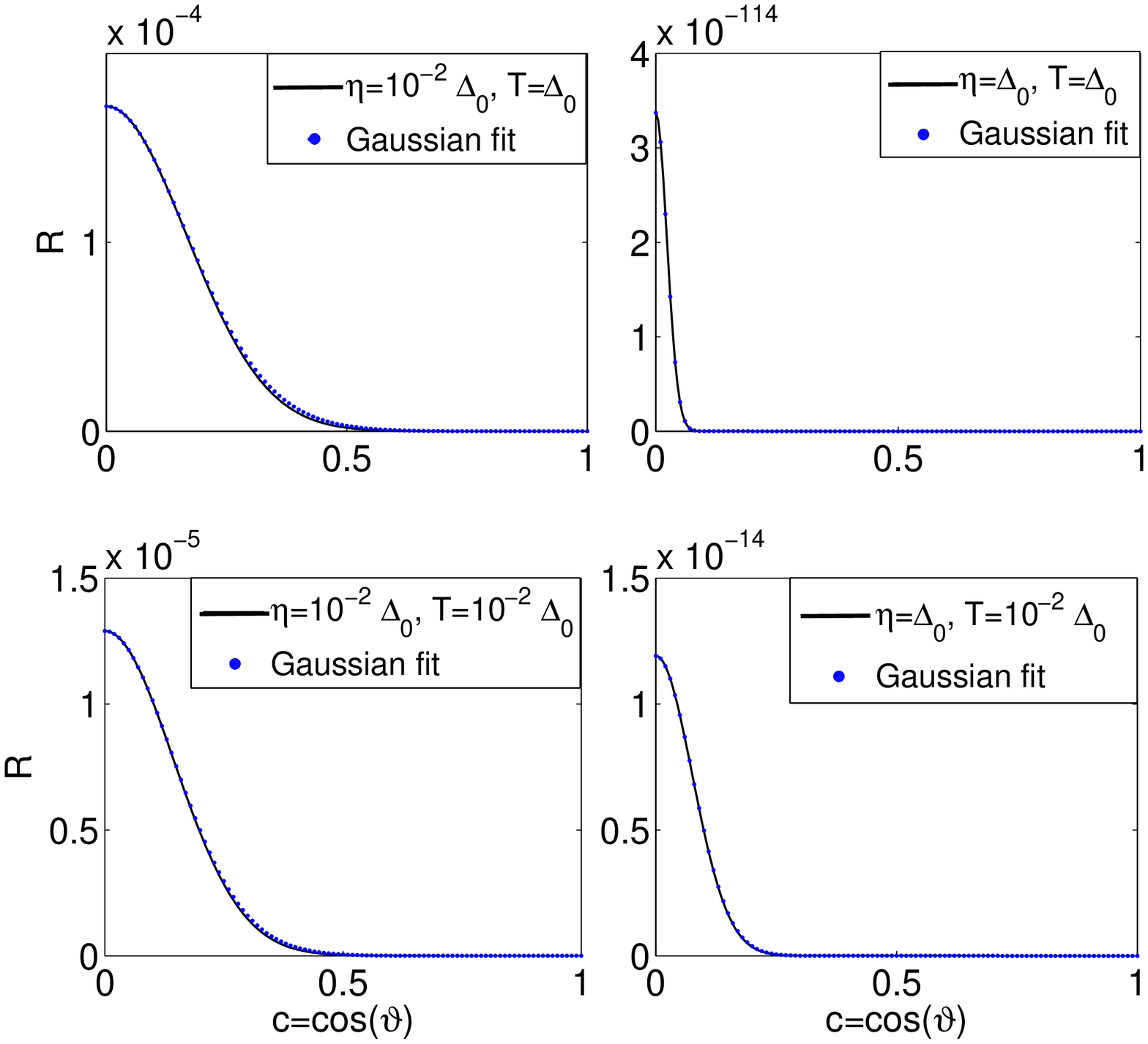}
\caption[]{Reduction factor from Equation (\ref{eq:reduction}) for superfluidity type B for different values of $\eta$ and $T$ as a function of the angle. The blue dots are the reduction factor calculated with the trapezoidal rule and the solid line is our analytical approximation to the reduction factor.}
\label{fig:factor_tipoB}
\end{figure*} 
\label{sec:appendix_1}
To numerically deal with the anisotropy in the reduction factors, it is necessary to integrate over the angle $\vartheta$ (see section \ref{neutrino-section}). For this purpose, we evaluate and inspect the shape of the integrand of the reduction factor as a function of the angle. The behaviour of this factor varies strongly with the temperature and the chemical imbalance, thus a range of values needs to be considered. Figure \ref{fig:factor_tipoB} shows this analysis for type B superfluidity, from which we conclude that the integrand can be approximated by a Gaussian function,
\begin{eqnarray}
\label{eq:R-1}
R=\int_0^{\pi/2}a\exp(-b\cos^2\vartheta) d\vartheta,
\end{eqnarray} 
with $a$ and $b$ constants, to better than 0.3\% (for every combination of $\Delta$ and $\eta$). We calculate this, evaluating Equations (\ref{eq:murca_inicial}) and (\ref{eq:murca_inicial2}) for only two different angles, in order to find the values of $a$ and $b$, this way calculating the integral, allowing us to increase the calculation time by only one evaluation in every reduction factor calculated relative to the isotropic case.

\begin{figure*}
\includegraphics[width=0.65\textwidth]{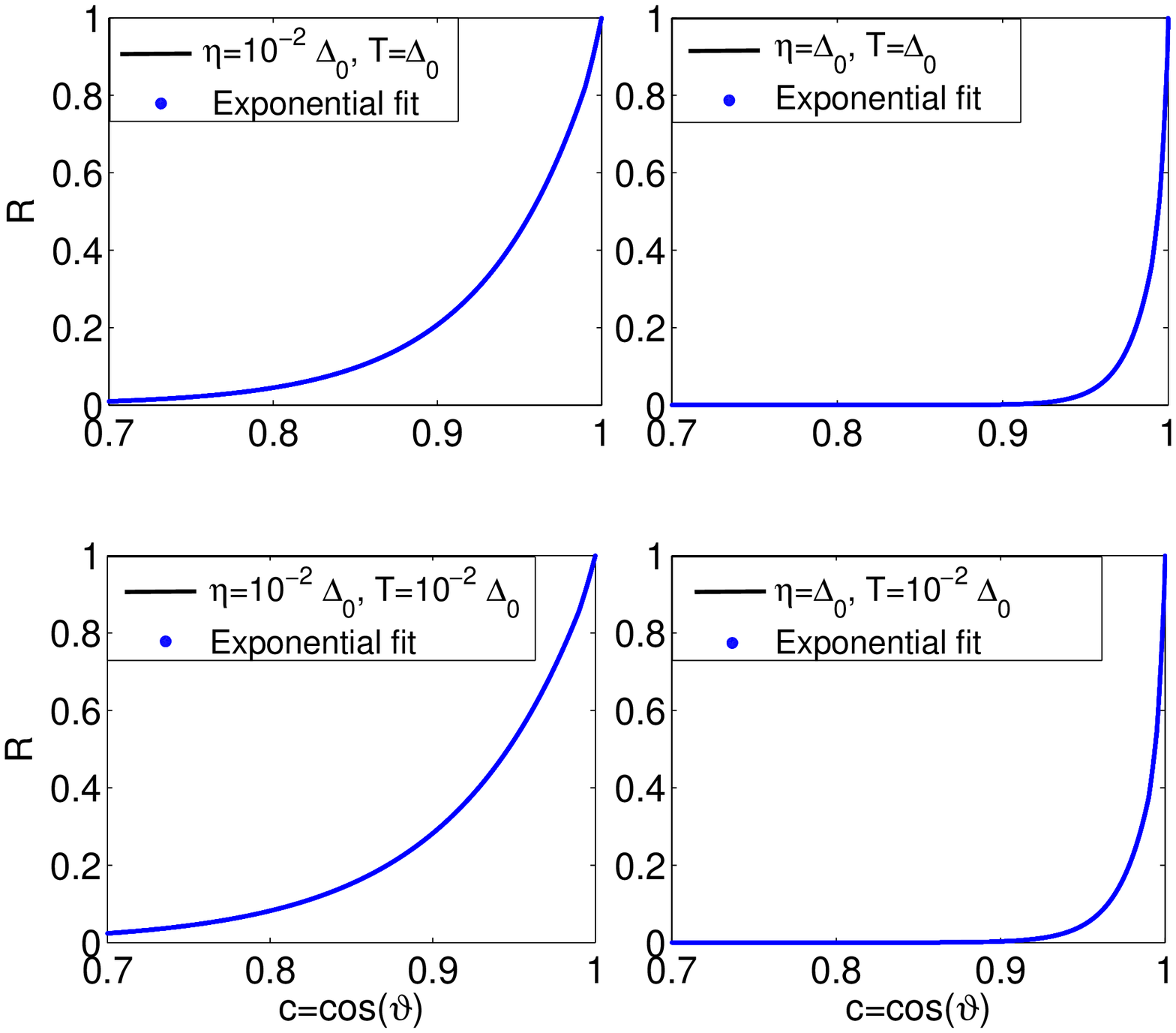}
\caption[]{Reduction factor of eq. \ref{eq:reduction} for superfluidity type C for different values of $\eta$ and $T$ as a function of the angle. The blue dots are the reduction factor calculated with the trapezoidal rule and the solid line is our analytical approximation to the reduction factor.}
\label{fig:factor_tipoC}
\end{figure*}

Figure \ref{fig:factor_tipoC} shows the behaviour of the type C superfluidity. From this we conclude that an acceptable approximation (with less than 1.5\% error) is a combination of exponential functions of the form:

\begin{eqnarray}
 R=\sum_{j=0}^{6}\int_{\displaystyle c_j}^{\displaystyle c_{j+1}}\exp^{\displaystyle(a_j+b_j\cos\vartheta)}d\vartheta
\end{eqnarray}
with $a_j, b_j, c_j$ constants. In order to find the values of $a_j$ and $b_j$ we need to evaluate the equations for seven angles, and this implies a large cost in calculation time compared with type B superfluidity, but in any case this approach considerably reduces the time relative to the direct use of  $F(\vartheta)$. To do the integration, we use $c=\cos(\vartheta)$; then the integration is over the range $0\leq c\leq 1$, and $c_j=0,0.7,0.95,0.99,0.995,0.997,0.9999,1$. This discretization is based on our identification of the points where the integrand changes significantly.


\begin{thebibliography}{}

\bibitem[Akmal et al.(1998)]{apr98} Akmal, A., Pandharipande, V. R. \& Ravenhall, D. G., 1998, Phys. Rev. C, 58, 1804

\bibitem[Alpar et al.(1984)]{alpar} Alpar, M. A., Anderson, P. W., Pines, D., \& Shaham, J. 1984, ApJ, 276, 325

\bibitem[Amundsen \& \O{stgaard} (1985a,b)]{admun85} Amundsen, L. \& \O{stgaard}, E., 1985a, Nuclear Physics A, 437, 2, 487.

\bibitem[Amundsen \& \O{stgaard} (1985b)]{admun85b} Amundsen, L. \& \O{stgaard}, E., 1985b, Nuclear Physics A, 442, 1, 163.

\bibitem[Andersson et al.(2005)]{anderson} Andersson, N., Comer, G. L. \& Glampedakis, K., 2005, Nuclear Physics A, 763, 212

\bibitem[Antoniadis et al.(2013)]{anto13} Antoniadis J. et al., 2013, Science, 340, 448


\bibitem[Baldo et al.(1998)]{modelHIJ}  Baldo, M., Elgar\o y, \O., Engvik,L., Hjorth-Jensen, M. \& Schulze,H.J., 1998, Phys. Rev C 58 1921

\bibitem[Bardeen, Cooper \& Schrieffer(1957)]{BCS} Bardeen, J., Cooper, L. \& Schrieffer, J., 1957, Phys. Rev., 108, 1175

\bibitem[Bogdanov (2013)]{bog13} Bogdanov, S., 2013, ApJ., 762, 96

\bibitem[Demorest et al.(2010)]{demo}Demorest, P. B., Pennucci, T., Ransom, S. M. Roberts, M. S. E. \& Hessels, J. W. T., 2010, Nature, 467, 7319, 1081

\bibitem[Durant et al.(2012)]{Durant} Durant, M., Kargaltsev, O., Pavlov, G., Kowalski, P. M., Posselt, B., van Kerkwijk, M. H. \& Kaplan, D. L., 2012, ApJ, 746, 1, 6.

\bibitem[Elgar\o y et al.(1996)]{modelKL} Elgar\o y, \O ., L. Engvik, M. Hjorth-Jensen \& E. Osnes, 1996, Nucl. Phys A 607 425

\bibitem[Elgar\o y et al.(1996b)]{modelM} Elgar\o y, \O ., L. Engvik, M. Hjorth-Jensen \& E. Osnes, 1996, Phys. Rev. Lett. 77 1428

\bibitem[Elshamouty et al.(2013)]{elsha} Elshamouty, K. G., Heinke, C. O., Sivakoff, G. R., et al. 2013, ApJ, 777, 22

\bibitem[Fern\'andez \& Reisenegger (2005)]{F-R} Fern\'andez, R. \& Reisenegger, A. 2005, ApJ, 625, 291

\bibitem[Flowers et al.(1976)]{flowers} Flowers, E., Ruderman, M. \& Sutherland, P., 1976, ApJ., 205, 541

\bibitem[Glendenning(1997)]{glend97} Glendenning, N. K., 1997, Compact Stars (Springer)

\bibitem[Gonz\'alez \& Reisenegger(2010)]{denix} Gonz\'alez, D. \& Reisenegger A., 2010, A\&A, 522, A16

\bibitem[Gonz\'alez-Caniulef \& Reisenegger(2010)]{denix2} Gonz\'alez-Caniulef, D. \& Reisenegger A., 2014, ApJ, submitted

\bibitem[Gudmundsson et al.(1983)]{gpe83} Gudmundsson, E. H., Pethick, C. J. \& Epstein, R. I., 1983, ApJ, 272, 286


\bibitem[Haensel(1992)]{haen92} Haensel, P., 1992, A\&A, 262, 131

\bibitem[Hebeler et al.(2013)]{heb13} Hebeler, K., Lattimer, J. M., Pethick, C. J., \& Schwenk, A., 2013, ApJ, 773, 11

\bibitem[Heinke \& Ho (2010)]{heinke} Heinke, C. O., \& Ho, W. C., 2010, ApJ, 719, L167



\bibitem[Kaminker et al.(2001)]{kam01}Kaminker, A. D., Haensel, P. \& Yakovlev, D. G., 2001, A\&A, 373, L17

\bibitem[Kargaltsev et al.(2004) Kargaltsev, Pavlov \& Romani]{kargaltsev04} Kargaltsev, O., Pavlov, G. G. \& Romani, R., 2004, ApJ, 602, 327

\bibitem[Levenfish \& Yakovlev(1994)]{levyak94} Levenfish, K. P. \& Yakovlev, D. G., 1994, ARep, 38, 247

\bibitem[Lombardo \& Schulze(2001)]{lombardo} Lombardo, U. \& Schulze, H., 2001,  Lecture Notes in Physics, 578, 30

\bibitem[Lyne et al (2004)]{lyne} Lyne, A. G., et al., 2004, Science, 303, 1153

\bibitem[Migdal(1960)]{migdal60} Migdal, A. B., 1960, Soviet Physics JETP 10, 176

\bibitem[Page et al.(2011)]{page11}Page, D., Prakash, M., Lattimer, J. M., \& Steiner, A. W., 2011, Phys. Rev. Lett.,106, 081101

\bibitem[Petrovich \& Reisenegger(2010)]{petroka} Petrovich, R. \& Reisenegger, A., 2010, A\&A, 521, A77 (PR10)

\bibitem[Petrovich \& Reisenegger(2011)]{petro11} Petrovich, R. \& Reisenegger, A., 2011, A\&A, 528, A66 

\bibitem[Posselt et al. (2013)]{posselt} Posselt, B., Pavlov, G. G., Suleimanov, V. \& Kargaltsev, O., 2013, ApJ, 779, 186

\bibitem[Potekhin et al.(1997)]{potek} Potekhin, A. Y., Chabrier, G., \& Yakovlev, D. G., 1997, A\&A, 323, 415

\bibitem[Reisenegger(1995)]{reis95} Reisenegger, A., 1995, ApJ, 442, 749

\bibitem[Reisenegger(1997)]{reis97} Reisenegger, A., 1997, ApJ, 485, 313

\bibitem[Reisenegger et al.(2006)]{reis06} Reisenegger, A., Jofr\'e, P., Fern\'andez, R. \&  Kantor, E., 2006, ApJ, 653, 568

\bibitem[Shternin et al.(2011)]{Shternin} Shternin, Peter S., Yakovlev, Dmitry G., Heinke, Craig O., Ho, Wynn C. G., Patnaude, \& Daniel J., 2011, MNRAS,  412, 1, L108


\bibitem[Thorne(1977)]{thorne77} Thorne, K. S., 1977, ApJ, 212, 825


\bibitem[Verbiest et al.(2008)]{verbiest} Verbiest, J., et al., 2008, ApJ, 679, 675

\bibitem[Yakovlev et al.(1999) ]{yak99} Yakovlev, D.G., Levenfish, K.P., Shibanov, Yu.A.,  1999, Phys.Usp., 42, 737
\bibitem[Yakovlev et al.(2001) ]{yak01} Yakovlev, D. G., Kaminker, A.D., Gnedin, O.Y. \& Haensel, P., 2001, Phys, Rep, 354, 1
\bibitem[Yakovlev et al.(2004) ]{yak04} Yakovlev, D. G., Gnedin, O. Y., Kaminker, A. D., Levenfish, K. P. \& Potekhin, A. Y., 2004, AdSpR, 33, 523
\bibitem[Yakovlev et al.(2008)]{yak08} Yakovlev D. G., 2008, in Bassa C., Wang Z., Cumming A. \& Kaspi V., 2008, eds., AIP Conf. Proc. V. 983. 40 Years of Pulsars: Millisecond Pulsars, Magnetars and More. Am. Inst. Phys., Melville, NY, 379


\end{thebibliography}
\end{document}